\documentclass[aps,prd,amsmath,amssymb,preprintnumbers,nofootinbib,a4paper,11pt]{revtex4-1}
\pdfoutput=1
\usepackage{amsthm}
\usepackage{graphicx}
	\graphicspath{{./NewFig/}}
\usepackage{url}
\usepackage[bookmarks, pagebackref=false]{hyperref}
\usepackage{color}
	\definecolor{rossoCP3}{cmyk}{0,.88,.77,.40}
		\definecolor{graa}{rgb}{0.8,0.8,0.8}
		\definecolor{blaa}{rgb}{0.2,0.2,0.6}
		\hypersetup{
			colorlinks,
			bookmarksopen,
			bookmarksnumbered,
			citecolor=blaa, 		
			linkcolor=rossoCP3,	
			urlcolor=rossoCP3,			
			}
\usepackage{dcolumn}
\usepackage{bm}
\usepackage{bbm}
\usepackage{pxfonts}

\usepackage{epsfig}
\usepackage{placeins}

\usepackage[ margin=5pt, font=small,labelfont=bf, justification=raggedright]{caption}
\usepackage{youngtab}
\usepackage{slashed}
\usepackage{subfig}

\newcommand{\beq}{\begin{eqnarray}}
\newcommand{\eeq}{\end{eqnarray}}

\newcommand{\bmp}{\noindent\begin{minipage}{16cm}}
\newcommand{\emp}{\end{minipage}\vskip 7mm} 

\def\lsim{\mathrel{\rlap{\lower4pt\hbox{\hskip1pt$\sim$}}
    \raise1pt\hbox{$<$}}}                
\def\gsim{\mathrel{\rlap{\lower4pt\hbox{\hskip1pt$\sim$}}
    \raise1pt\hbox{$>$}}}                

\begin{document}

\title{\Large  \color{rossoCP3}  Light Higgs from Scalar See-Saw in Technicolor
}
\author{Roshan Foadi}
\email{roshan.foadi@uclouvain.be}
\affiliation{Centre for Cosmology, Particle Physics and Phenomenology (CP3)
Chemin du Cyclotron 2, Universit\'e catholique de Louvain, Belgium}
\author{Mads T. {Frandsen}}
\email{m.frandsen@cp3-origins.net}
\affiliation{
{ CP}$^{ \bf 3}${-Origins} and the Danish Institute for Advanced Study \\
University of Southern Denmark, Campusvej 55, DK-5230 Odense M, Denmark.}


\begin{abstract}
We consider a TeV scale see-saw mechanism leading to light scalar resonances in models with otherwise intrinsically heavy scalars. The mechanism can provide a 125 GeV technicolor Higgs in {\em e.g.} two-scale TC models. \\
[.5cm]
{
\small \it {Preprint: CP$\,^3$-Origins-2012-32 \& DIAS-2012-33}}
\end{abstract}

\maketitle
\newpage
\section{Introduction}
Recently the ATLAS and CMS collaborations have announced the discovery of a new boson with a mass of approximately 125 GeV~\cite{:2012gk,:2012gu}. The next challenge is to determine the nature of this new state, including its quantum numbers and couplings, and whether it is fundamental or composite. Because of the observed decays to $\gamma\gamma$, $WW$, and $ZZ$, with strengths seemingly comparable to those expected from the standard model (SM) Higgs, this state is likely a (mostly) $CP$-even spin-zero boson, although $CP$-odd candidates are not currently ruled out~\cite{Frandsen:2012rj,Barroso:2012wz,Coleppa:2012eh,Chivukula:2012cp,Eichten:2012qb}.

In technicolor (TC), {\it common lore} has it that the lightest $CP$-even spin-zero resonance, the analogue of the $\sigma$ meson or $f_0(500)$ in QCD, cannot be as light as 125 GeV. In fact TC theories are sometimes considered as underlying theories for Higgsless models, despite the fact that in QCD the $\sigma$ meson is among the lightest states. In this paper we consider the possibility of a light {\it TC Higgs} arising from mass mixing between relatively heavy scalar singlets. This results in a ``see-saw'' mechanism, with one scalar singlet becoming lighter and one heavier than the corresponding diagonal mass. This is expected to occur in two-scale TC models, {\it e.g.} low-scale TC~\cite{Lane:1989ej,Eichten:2011sh} and ultra-minimal TC (UMT)~\cite{Ryttov:2008xe}\footnote{In models with fundamental scalars, a scalar see-saw mechanism has also been considered as a way of generating a negative mass squared for the Higgs \cite{Calmet:2002rf,Calmet:2006hs}.}.

Two-scale TC theories feature two technifermion species with different representations under a single technicolor gauge group. These lead, for instance, to two different sets of composite scalars. Because of different quantum numbers, scalar multiplets from different representations do not mix through mass terms. However scalar singlets do. Because of the strength of the TC interaction, such a mixing can be sizable, which is the key ingredient in the see-saw mechanism. Moreover, radiative corrections from the top quark may contribute to further reduce the mass of the lightest scalar singlet~\cite{Foadi:2012bb}.

This paper is organized as follows. In Sec.~\ref{sec:seesaw} we briefly review the see-saw mechanism for scalar singlets. In Sec.~\ref{sec:twoscale} we review the spin-zero sector of two-scale TC, and analyze the properties of the mixing mass term. Then we apply the general results to UMT and low-scale TC. 
Finally, in Sec.~\ref{sec:conclusions} we offer a brief discussion of our findings.
\section{Higgs see-saw mechanism}\label{sec:seesaw}
\begin{figure}[t!]
\begin{center}
 \includegraphics[width=0.55\linewidth]{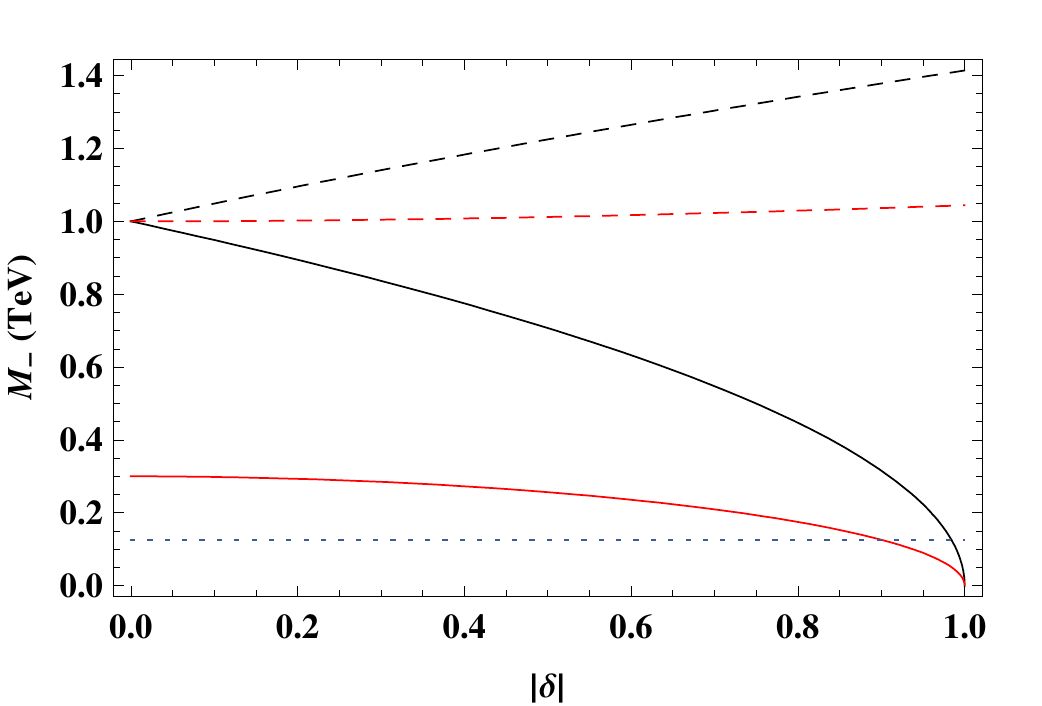}
\caption{$M_-$ (solid) and $M_+$ (dashed) as a function of $|\delta|$, for the cases $M_1=M_2=1.0$ TeV (black) and $M_1=1.0$ TeV, $M_2=300$ GeV (red). The dotted horizontal line corresponds to the experimental value of 125 GeV.}
\label{fig:eigenvalues}
\end{center}
\end{figure}

Consider a theory featuring two scalar singlets in its spectrum, $H_1$ and $H_2$. Assume these to mix via mass term:
\begin{eqnarray}
\mathcal{L}\supset -\frac{M_1^2}{2}H_1^2 - \frac{M_2^2}{2} H_2^2  - \delta\ M_1 M_2\ H_1 H_2 \ .
 \label{Eq:basepotential}
\end{eqnarray}
In the limit $\delta^2\to 1$ one eigenstate is massless. It is therefore useful to define the parameter
\begin{equation}
\varepsilon\equiv 1-\delta^2 \ .
\end{equation}
Diagonalization gives
\begin{equation}
\left(\begin{array}{c} H_1 \\ H_2 \end{array}\right) = \left(\begin{array}{cc} \cos\beta & \sin\beta \\ -\sin\beta & \cos\beta \end{array}\right)
\left(\begin{array}{c} H_- \\ H_+ \end{array}\right)\ , \quad
\tan2\beta = \frac{2 M_1 M_2}{M_2^2-M_1^2}\delta \ ,
\end{equation}
where $H_-$ and $H_+$ are the light and heavy mass eigenstate, respectively, with mass
\begin{equation}
M_\pm^2 = \frac{M_1^2+M_2^2}{2}\left[1\pm\sqrt{1-\left(\frac{2 M_1 M_2}{M_1^2+M_2^2}\right)^2\varepsilon}\right] \ .
\end{equation}
For $\varepsilon\ll 1$, $M_-^2$ becomes
\begin{equation}
M_-^2 = \frac{M_1^2 M_2^2}{M_1^2+M_2^2}\varepsilon + {\cal O}(\epsilon^2) \ll M_1^2,\ M_2^2 \ .
\end{equation}
In Fig.~\ref{fig:eigenvalues} we show $M_-$ (solid) and $M_+$ (dashed), for the cases $M_1=M_2=1.0$ TeV (black) and $M_1=1.0$ TeV, $M_2=300$ GeV (red), as a function of $|\delta|$. The dotted horizontal line corresponds to the experimental value of 125 GeV. This trivial exercise illustrates how theories with relatively heavy scalar mass scales, {\it e.g.} $M_i$ between a few hundreds of GeVs and a few TeVs, may still feature a light scalar eigenstate after diagonalization, and thus a candidate for the recently observed 125 GeV boson. While this mechanism is simple and general,
it is immediately applicable to TC models with two condensation scales. In the next section we discuss the general properties of singlet scalar mixing in two-scale TC theories, and provide specific examples.
\section{Two-scale technicolor}\label{sec:twoscale}
In two-scale TC two dynamical scales arise due to the presence of two Dirac technifermion species $Q_i$, transforming under different representations $R_i$ of a single TC gauge group~\cite{Lane:1989ej,Ryttov:2009yw}. The TC force causes technifermion bilinears to condense at different scales $\Lambda_i$. An estimate of $\Lambda_2/\Lambda_1$ can be obtained from the ladder Schwinger-Dyson equation for the techniquark propagator.  In this approximation, the critical coupling for chiral symmetry breaking depends on the representation via $\alpha_c (R_i)=\pi/3 C_2(R_i)$, where $C_2(R_i)$ is the Casimir of the representation $R_i$. Taking $C_2(R_1)\leq C_2(R_2)$, and integrating the one-loop beta-function $\beta(\alpha) = -\beta_0(R) \alpha^2/2\pi$ from $\Lambda_1$ to $\Lambda_2$ gives
\begin{equation}
\frac{\Lambda_2}{\Lambda_1} \simeq \exp\left[\frac{2\pi}{\beta_0(R_1)}\bigg(\alpha_\mathrm{c}(R_2)^{-1}-\alpha_\mathrm{c}(R_1)^{-1} \bigg)\right]\ .
\label{Eq:scaleratio}
\end{equation}
Since $\Lambda_1 \leq \Lambda_2$, or equivalently $\alpha_c(R_1) \geq\alpha_c(R_2)$, the fermions in the representation $R_2$ are effectively decoupled below $\Lambda_2$. Therefore, only $\beta_0(R_1)$ appears in the exponent. If $\beta_0(R_1)$ and $\alpha_c(R_1)$ are
small then the scale separation can be sizeable and the presence of four-fermion operators can contribute to further enhance the scale separation~\cite{Frandsen:2011kt}. This crude approximation serves to illustrate the appearance of two distinct scales.

Now, let $N_i$ be the number of Dirac techniflavors in the representation $R_i$. The global symmetries of the corresponding fermion sector depend on whether $R_i$ is complex, real, or pseudoreal . To see this we express the Dirac fermions in terms of two Weyl fermions,
\begin{equation}
Q_{im} = \left(\begin{array}{c}\psi^1_{im} \\ \bar{\psi}^2_{im}\end{array}\right)\ ,
\end{equation}
where $m$ is the techniflavor index ($m=1,\dots, N_i$), and the color index is suppressed. If $\psi^1_{im}$ transforms under the $R_i$ representation, then $\psi^2_{im}$ transforms under the conjugate representation $\overline{R}_i$. For complex $R_i$ this implies that rotations in techniflavor space {\it cannot} mix $\psi^1$ and $\psi^2$ fermions. As a consequence the TC Lagrangian features a global $SU(N_i)_1\times SU(N_i)_2\times U(1)$ techniflavor symmetry, where the extra $U(1)$ corresponds to technibaryon-number conservation. This global symmetry is spontaneously broken to diagonal $SU(N_i)\times U(1)$ by the condensate. The latter is an $N_i\times N_i$ complex matrix,
\begin{equation}
\left(\phi_i\right)_{mn} \sim \psi^1_{im} \psi^2_{in} \ ,
\end{equation}
which transforms like the bi-fundamental of $SU(N_i)_1\times SU(N_i)_2$:
\begin{equation}
\phi_i\to u_{i1} \phi_i u_{i2}^\dagger\ ,\quad u_{iA}\in SU(N_i)_A \ .
\end{equation}
In terms of spin-zero composites $\phi_i$ reads
\begin{equation}
\phi_i =
\frac{v_i + H_i + i\Theta_i}{\sqrt{2N_i}} + \left(i \Pi_i^a + \Sigma_i^a\right)T_i^a\ ,
\end{equation}
where $T_i^a$ are the $SU(N_i)$ broken generators, normalized according to ${\rm Tr}\ T_i^a T_i^b = \delta^{ab}/2$. Here $v_i$ is the vacuum expectation value of the condensate, and $H_i$ is a $\psi_{im}^1\psi^2_{im}$ scalar singlet.

If the representation $R_i$ is real, rotations in techniflavor space {\it can} mix $\psi^1$ and $\psi^2$ fermions, as these transform in the same way under the TC gauge group. As a consequence the TC Lagrangian features a global $SU(2N_i)$ techniflavor symmetry, which is spontaneously broken to $SO(2N_i)$ by the condensate. The latter is a $2N_i\times 2N_i$ complex matrix,
\begin{equation}
\left(\Phi_i\right)_{mn}^{AB} \sim \psi^A_{im} \psi^B_{in} \ ,
\label{eq:condensateReal}
\end{equation}
where $A,B=1,2$, and $m,n=1,\dots, N_i$. The spin-zero matrix $\Phi_i$ transforms as the two-index symmetric representation of $SU(2N_i)$:
\begin{equation}
\Phi_i\to u_i \Phi_i u_i^T\ ,\quad u_i\in SU(2N_i)\ ,\quad \left(\Phi_i\right)_{nm}^{BA} = \left(\Phi_i\right)_{mn}^{AB}\ .
\end{equation}
In terms of spin-zero composites, $\Phi_i$ reads
\begin{equation}
\Phi_i = \left[\frac{v_i + H_i + i\Theta_i}{\sqrt{4N_i}} + \left(i \Pi_i^a + \Sigma_i^a\right)X_i^a\right] E_i\ ,
\label{eq:2N2N}
\end{equation}
where $X_i^a$ are the broken generators belonging to the $SU(2N_i)-SO(2N_i)$ algebra, and normalized according to ${\rm Tr}\ X_i^a X_i^b = \delta^{ab}/2$. The matrix $E_i$ is an $SO(2N_i)$ invariant satisfying
\begin{equation}
E_i^T {X_i^a}^T = X_i^a E_i\ ,\quad E_i^T = E_i\ ,\quad E_i E_i^\dagger = 1\ .
\end{equation}

Finally, if the $R_i$ representation is pseudoreal, the global symmetry in techniflavor space is $SU(2N_i)$, and the condensate is as in Eq.~(\ref{eq:condensateReal}). However the matrix $\Phi_i$ is now in the two-index antisymmetric representation of $SU(2N_i)$,
\begin{equation}
\Phi_i\to u_i \Phi_i u_i^T\ ,\quad u_i\in SU(2N_i)\ ,\quad \left(\Phi_i\right)_{nm}^{BA} = - \left(\Phi_i\right)_{mn}^{AB}\ ,
\end{equation}
because of an extra minus sign introduced by the invariant which contracts the TC indices (not displayed in Eq.~(\ref{eq:condensateReal})). As a consequence the condensate breaks the global symmetry to $Sp(2N_i)$ rather than $SO(2N_i)$. In terms of spin-zero composites, the $2N_i\times 2N_i$ $\Phi_i$ matrix is as in Eq.~(\ref{eq:2N2N}), where now the $X_i^a$ broken generators belong to the $SU(2N_i)-Sp(2N_i)$ algebra, and $E_i$ is an $Sp(2N_i)$ invariant satisfying
\begin{equation}
E_i^T {X_i^a}^T = - X_i^a E_i\ ,\quad E_i^T = - E_i\ ,\quad E_i E_i^\dagger = 1\ .
\end{equation}

We can unify notation for the complex and real or pseudoreal scenarios by defining, for the case of complex $R_i$, the $2N_i\times 2N_i$ matrix
\begin{equation}
\Phi_i \equiv \frac{1}{\sqrt2} \left(\begin{array}{cc} 0 & \phi_i \\ \phi_i^\dagger & 0 \end{array}\right) \ .
\end{equation}
Then, assuming no violation of techniflavor symmetry, and retaining only terms up to dimension four, the symmetry-breaking potential reads
\begin{eqnarray}
V &=& - \mu_1^2\ {\rm Tr}\ \Phi_1 \Phi_1^\dagger - \mu_2^2\ {\rm Tr}\ \Phi_2 \Phi_2^\dagger
+\lambda_1^\prime {\rm Tr}\ \Phi_1 \Phi_1^\dagger \Phi_1 \Phi_1^\dagger
+\lambda_1^{\prime\prime} {\rm Tr}\ \Phi_1 \Phi_1^\dagger\ {\rm Tr}\ \Phi_1 \Phi_1^\dagger \nonumber \\
&+& \lambda_2^\prime {\rm Tr}\ \Phi_2 \Phi_2^\dagger \Phi_2 \Phi_2^\dagger
+\lambda_2^{\prime\prime} {\rm Tr}\ \Phi_2 \Phi_2^\dagger\ {\rm Tr}\ \Phi_2 \Phi_2^\dagger
+ 2\lambda {\rm Tr}\ \Phi_1 \Phi_1^\dagger\ {\rm Tr}\ \Phi_2 \Phi_2^\dagger \ .
\label{Eq:twoscalepot}
\end{eqnarray}
A few comments are in order for this potential. First, the pseudoscalars are all massless in Eq.~(\ref{Eq:twoscalepot}). In particular, the $\Pi_i^a$ fields are the Nambu-Goldstone bosons (NGBs) associated to the spontaneous breaking of the $SU(N_i)_1\times SU(N_i)_2$ or $SU(2N_i)$ techniflavor symmetries. Three of these NGBs become the longitudinal components of the SM $W$ and $Z$ boson, once the electroweak interactions are ``switched on''. The remaining NGBs receive mass through radiative effects and/or additional new interactions beyond TC, such as Extended TC \cite{Eichten:1979ah,Dimopoulos:1979es}. These interactions can be accounted for by adding techniflavor-breaking potential terms to Eq.~(\ref{Eq:twoscalepot}). The $\Theta_i$ pseudoscalar singlets are massless in Eq.~(\ref{Eq:twoscalepot}), because of additional and spontaneously broken $U(1)_i$ symmetries. These states acquire mass from instantons -- in the form of ${\rm Det}\ \Phi_i$ invariants to be added to $V$ -- and/or ETC interactions. Finally, the scalar multiplets $\Sigma_i^a$ can be made heavier than the singlets (as expected from scaling up the QCD spectrum) by adjusting the quartic terms, and/or by including higher order invariant terms to the potential. We shall ignore all these issues, as our goal is to highlight the see-saw mechanism for the scalar singlets. This is fully accounted for in the potential of Eq.~(\ref{Eq:twoscalepot}).

Minimization of the potential gives
\begin{equation}
v_1^2 = \displaystyle{\frac{\mu_1^2/\lambda_1-\lambda\mu_2^2/\lambda_1\lambda_2}{1-\lambda^2/\lambda_1\lambda_2}} \ ,\quad
v_2^2 = \displaystyle{\frac{\mu_2^2/\lambda_2-\lambda\mu_1^2/\lambda_1\lambda_2}{1-\lambda^2/\lambda_1\lambda_2}}\ ,
\end{equation}
where
\begin{equation}
\lambda_i\equiv \frac{\lambda_i^\prime}{2 N_i} + \lambda_i^{\prime\prime}\ .
\end{equation}
The isosinglet mass Lagrangian is as in Eq.~(\ref{Eq:basepotential}), with
\begin{equation}
M_i^2 = 2\lambda_i v_i^2\ ,
\end{equation}
and
\begin{equation}
\delta^2 = \frac{\lambda^2}{\lambda_1\lambda_2}\ .
\end{equation}

\begin{figure}[t!]
\begin{center}
\includegraphics[width=10.0cm]{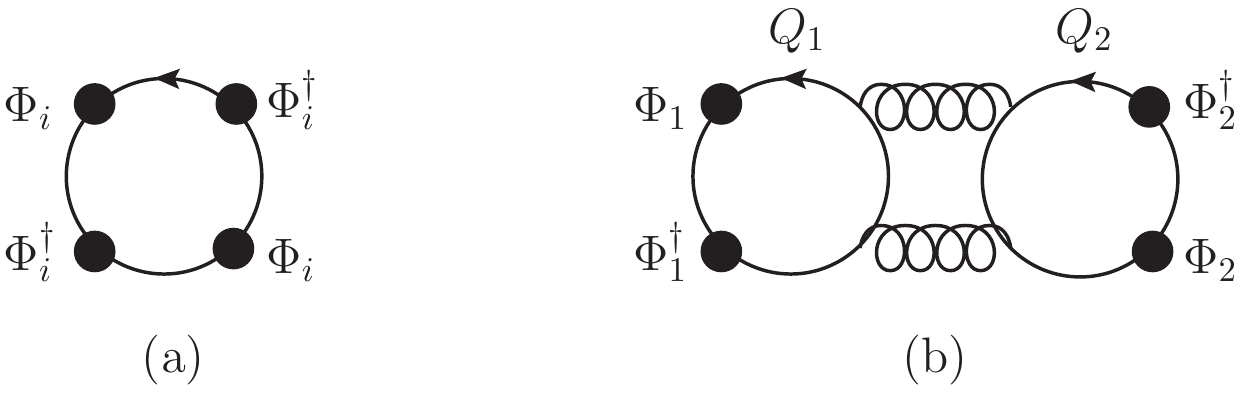}
\end{center}
\caption{}
\label{Fig:diagrams}
\end{figure}

In an $SU(N_{\rm TC})$ theory, in the limit of large
$N_{\rm TC}$, the double trace terms, in particular the mass mixing term, are subleading in $1/N_{\rm TC}$, see {\it e.g.} \cite{Manohar:1998xv}. To see this explicitly, consider that the contributions to $\lambda_i$ and $\lambda$ are dominated by the diagrams of Fig.~\ref{Fig:diagrams} (a) and (b), respectively, where the black disks represent scalar insertions. These introduce a normalization factor $1/\sqrt{d(R_i)}$, where $d(R_i)$ is the dimension of the representation $R_i$. Recalling that the TC gauge coupling scales like $1/\sqrt{N_{\rm TC}}$, we find the scaling behaviors
\begin{equation}
\lambda_i \sim \frac{1}{N_i d(R_i)}\ ,\quad
\lambda\sim \frac{T(R_1)T(R_2)d(G)}{d(R_1)d(R_2)N_{\rm TC}^2}\ ,
\end{equation}
Here $T(R_i)$ is defined by ${\rm Tr}\ t_i^a t_i^b=T(R_i)\delta^{ab}$, where $t_i^a$ are the TC generators in the representation $R_i$, and $d(G)$ is the dimension of the TC group, $N_{\rm TC}^2-1$. In the large-$N_{\rm TC}$ limit this gives
\begin{equation}
\delta^2\sim\frac{N_1 N_2 T(R_1)^2 T(R_2)^2}{d(R_1) d(R_2)}\ ,
\label{eq:deltascaling}
\end{equation}
For example, if $R_1$ is the fundamental and $R_2$ the adjoint representation, then $\delta^2\sim N_1 N_2/N_{\rm TC}$. The fact that $\delta^2$ decreases with $N_{\rm TC}$ was to be expected, as $\lambda$ arises from a three-loop diagram, and is therefore subdominant in the large-$N_{\rm TC}$ limit. However, for small values of $N_{\rm TC}$ we expect $\delta^2$ to be of order one, as suppression from loop factors is compensated by the large TC coupling. From Eq.~(\ref{eq:deltascaling}) we also observe that $\delta^2$ grows with the number of flavors.

In addition to mass mixing of scalar singlets, we also expect mass mixing of the pseudoscalar and spin-one singlets. These, however, may feature larger diagonal masses than those of the scalar singlets
We will not further address this point here.

\subsection{Ultra-minimal technicolor}
The UMT model~\cite{Ryttov:2008xe} is a two-scale TC model based on an $SU(2)_{\rm TC}$ gauge theory, with $N_1=2$ Dirac techniflavors in the fundamental representation, $U$ and $D$, and $N_2=1$ Dirac techniflavor in the adjoint representation, $\lambda$. The two fundamental technifermions are arranged in a doublet with respect to the weak interactions, whereas the adjoint technifermion is not charged under the electroweak interactions.
UMT features the smallest contribution to the {\it perturbative}~\footnote{By $S_{\rm pert}$ we mean the computation of $S$ from a loop of technifermions $Q$ with a dynamical mass $m_Q \gg m_Z$, see {\it e.g.} the discussion in \cite{Dietrich:2005jn}.} electroweak $S$ parameter, $S_{\rm pert}\simeq 1/3\pi$, compatible with electroweak symmetry breaking and near-conformal dynamics~\cite{Ryttov:2008xe,Ryttov:2009yw}.

In UMT both fermion species are assumed to condense at roughly the same scale $\Lambda_1 \simeq \Lambda_2 $. In fact it is readily found that $C_{2}(R_1)\simeq C_{2}(R_2)$, whence $\alpha_c (R_1) \simeq \alpha_c (R_2)$ in the ladder approximation. The technifermion-condensate vevs are $\langle \overline{U}_R U_L + \overline{D}_R D_L \rangle$ and $\langle \lambda^1 \lambda^2 \rangle$. The former breaks the electroweak symmetry and produces, among others, an isospin triplet of Goldstone bosons, $\Pi_1^{1,2,3}$. These become the longitudinal modes of the $W$ and $Z$ boson, which requires $v_1 = v=246$ GeV. Based on the above assumption, we also have $v_2 \simeq v_1$. The full global symmetry breaking pattern, in the absence of electroweak interactions, is $SU(4)\times SU(2) \times U(1) \to Sp(4) \times U(1) \times Z_2$. Notice the extra $U(1)$ symmetry, relative to the general symmetry breaking patterns discussed above. This arises from the fact that a linear combination of the two $U(1)$ symmetries, in $U(4)=SU(4)\times U(1)$ and $U(2)=SU(2)\times U(1)$, is anomaly free, whereas in isolation each one of these symmetries is anomalous.

Following the above discussion we can describe the scalar sector using a linear realization of the global symmetries in terms of a $4\times 4$ matrix $\Phi_1$, and a $2\times 2$ matrix $\Phi_2$. Up to dimension-four terms the potential is as in Eq.~(\ref{Eq:twoscalepot}). This gives nine massless pseudoscalars: $\Pi_1^{1,2,3,4,5}$ and $\Pi_2^{1,2}$, corresponding to $SU(4)\to Sp(4)$ and $SU(2)\to U(1)$ spontaneous symmetry breaking, respectively, plus the $\Theta_{1,2}$ pseudoscalar singlets. A linear combination of these is the Goldstone boson corresponding to the $U(1)\to Z_2$ spontaneous symmetry breaking, whereas the remaining linear combination receives mass from instantons, in the form of higher dimensional terms. The one of lowest order has dimension six:
\begin{eqnarray}
\left( \det \Phi_2 \right)^2 \text{Pf}\ \Phi_1 + \text{h.c.} \ .
\label{Eq:LUMT}
\end{eqnarray}
Incidentally this provides an additional mass mixing term for the two scalar singlets.  The UMT model is thus a prime TC example where the lightest scalar mass eigenstate can be as light as 125 GeV, as exemplified by the black curve in Fig.~\ref{fig:eigenvalues}.

Note however that before mass mixing the scalar masses in the UMT model might be well below the TeV scale due to the argued walking dynamics of the model. For example the mass of the lightest scalar in the UMT model has been estimated to be as low as 250 GeV in Ref.~\cite{Doff:2009na}.
This model computation does not take into account mass mixing between the scalars, but relies on the assumed near-conformality of the theory.
\subsection{Low-scale TC}
In the two-scale TC framework proposed early on in Ref.~\cite{Lane:1989ej} the dynamical assumption is that $\Lambda_1 \ll \Lambda_2$ and such models are also referred to as low-scale TC~\cite{Eichten:2011sh}. This hierarchy in scales requires choosing the representations such that
the quadratic Casimirs satisfy $C_2(R_1) \ll C_2(R_2)$ and/or lead to a small $\beta_0(R_1)$~\footnote{Achieving this typically requires a large number of technifermions in the representation $R_1$, such as the fundamental, or alternatively four-technifermion operators with large enough coefficients~\cite{Frandsen:2011kt}.}. The low-scale TC assumption that both sectors have the technifermions arranged in weak doublets implies that the electroweak scale $v$ must be related to $v_i$ via $v=\sqrt{N_1 v_1^2 + N_2 v_2^2}$ to ensure the correct $W$ and $Z$ masses. For non-large values of $N_1$, $v \simeq \sqrt{N_2} v_2$.

Again we can describe the scalar sector using a linear realization of the global symmetries in terms of appropriate matrices of composite fields $\Phi_i$. The diagonal mass $M_1$ is by construction relatively light \cite{Delgado:2010bb}, and mass mixing will further reduce its value. The corresponding scenario is similar to the one depicted by the red curves in Fig.~\ref{fig:eigenvalues}.
\section{Summary and discussion}\label{sec:conclusions}
In this paper we have discussed how a light composite scalar may arise in two-scale TC theories via mass mixing between relatively heavy scalar resonances. We have argued that the mass of the light scalar may be compatible with the recently observed $\sim 125$ GeV resonance and discussed a concrete minimal two-scale TC model, UMT~\cite{Ryttov:2009yw}, which provides mass mixing of the required order of magnitude. We have also discussed the mechanism in low-scale TC models~\cite{Lane:1989ej}, where the lightest scalar resonance is expected to be relatively light compared to the TeV scale, already before including effects of mass-mixing. Finally radiative corrections from the interaction with the top quark can further reduce the mass of the lightest scalar resonance~\cite{Foadi:2012bb}.

It will be interesting to study the phenomenology of TC models featuring a scalar see-saw mechanism in the light of LHC data, already indicating that the couplings of the scalar resonance to the SM fermions and gauge bosons must be SM Higgs-like.
\acknowledgements
We thank K. Schmidt-Hoberg and T. Ryttov for comments, discussions and suggestions.
The work of R.F. is supported by the Marie Curie IIF grant proposal 275012.
The work of M.T.F is supported by a Sapere Aude Grant.
\bibliography{HD}
\bibliographystyle{ArXiv}

\end{document}